\documentclass[prb,aps,twocolumn,showkeys,showpacs,amsmath,amsfonts,floatfix,superscriptaddress,nofootinbib]{revtex4-1} 
\usepackage{amssymb,amsmath,amstext}                
\usepackage{graphicx}                                               
\usepackage{tikz-network}
\usepackage{epstopdf}                                               
\usepackage{color}       
\usepackage{bm}
\usepackage{appendix}                                              
\usepackage[latin2]{inputenc}
\usepackage{ulem}
\usepackage{latexsym}
\usepackage{tabularx}
\usepackage{afterpage}
\usepackage{nonfloat}
\usepackage[colorlinks=true,citecolor=blue,linkcolor=magenta]{hyperref}
\usepackage{lipsum}

\def\be{\begin{equation}}
\def\ee{\end{equation}}
\def\bea{\begin{eqnarray}}
\def\eea{\end{eqnarray}}
\def\bi{\begin{itemize}}
\def\ei{\end{itemize}}

\newcommand{\jd}[1]{{\color{red}{{#1}}}}

\begin{document}

\title{ 
Bang-bang preparation of a quantum many-body ground state in a finite lattice:
optimization of the algorithm with a tensor network
}

\newcommand{\affilju}{
             Jagiellonian University, 
             Faculty of Physics, Astronomy and Applied Computer Science,
             Institute of Theoretical Physics, 
             ul. \L{}ojasiewicza 11, 30-348 Krak\'ow, Poland 
             }

\newcommand{\affilkac}{  
             Jagiellonian University, 
             Mark Kac Center for Complex Systems Research,
             ul. \L{}ojasiewicza 11, 30-348 Krak\'ow, Poland 
             }
             
\author{Ihor Sokolov}\affiliation{\affilju}  
\author{Jacek Dziarmaga}\affiliation{\affilju}\affiliation{\affilkac}

\date{April 16, 2025}

\begin{abstract}
A bang-bang (BB) algorithm prepares the ground state of a lattice quantum many-body Hamiltonian $H=H_1+H_2$ by evolving an initial product state alternating between $H_1$ and $H_2$. We optimize the algorithm with tensor networks in one and two dimensions. The optimization has two stages. In stage one, a shallow translationally-invariant circuit is optimized in an infinite lattice. In stage two, the infinite-lattice gate sequence is used as a starting point for a finite lattice where it remains optimal in the bulk. The prepared state requires optimization only at its boundary, within a healing length from lattice edges, and the gate sequence needs to be modified only within the causal cone of the boundary.
We test the procedure in the 1D and 2D quantum Ising model near its quantum critical point employing, respectively, the matrix product state (MPS) and the pair-entangled projected state (PEPS). At the boundary already the infinite-lattice sequence turns out to provide a more accurate state than in the bulk. 
\end{abstract}

\maketitle

\section{Introduction}
\label{sec:intro}

Understanding strongly correlated quantum many-body systems is a long-standing problem, especially in two spatial dimensions (2D). Exact diagonalization is limited to small system sizes, Monte Carlo approaches are hampered by the infamous sign problem, and tensor networks are limited by entanglement. In principle the entanglement is not a problem for quantum simulators/computers but the present-day noisy intermediate scale quantum (NISQ) devices \cite{Preskill2018quantumcomputingin} are reliable only for shallow circuits.

In this work we design the quantum approximate optimization algorithm (QAOA) \cite{farhi2014quantum} to prepare a ground state on a finite lattice. QAOA divides the target Hamiltonian into non-commuting terms, $H=H_1+H_2$, and after initialization in a product state performs a sequence of unitary evolutions alternating between $H_1$ and $H_2$. This bang-bang (BB) \cite{Cerezo2021,PhysRevX.7.021027} sequence of rotation angles is optimized to minimize the final energy in the target Hamiltonian $H$. A smaller number of BB steps, equal to the depth of the quantum circuit, is preferable.
The shallowness makes QAOA ideally suited for simulation with tensor networks, as demonstrated for infinite translationally-invariant systems in 1D \cite{BB_MPS} and 2D \cite{BB_iPEPS}. The classical simulation helps to prepare the ground state on a quantum hardware before it is subject to further quantum processing that goes beyond any classical simulation. 
In this work we consider finite lattices. The optimization is divided in two stages. In stage one, a shallow translationally-invariant circuit is optimized in an infinite system \cite{BB_MPS,BB_iPEPS}. In stage two, the infinite-lattice gate sequence is used as a starting point for a finite lattice where it remains optimal in the bulk. The prepared state requires optimization only at its boundary, within a healing length from lattice edges, and the gate sequence needs to be modified only within the causal cone of the boundary.

Typical ground states of quantum many-body systems can be represented efficiently by tensor networks~\cite{Verstraete_review_08,Orus_review_14,Nishino_review_2022} including the matrix product states in one dimension (1D)~\cite{fannes1992}, the projected entangled pair state (PEPS) in 2D~\cite{Nishino_2DvarTN_04,verstraete2004} and 3D~\cite{Vlaar2021}, or the multi-scale entanglement renormalization ansatz (MERA)~\cite{Vidal_MERA_07,Vidal_MERA_08,Evenbly_branchMERA_14,Evenbly_branchMERAarea_14}. 
In addition to the matrix product state in 1D, we employ the genuinely 2D PEPS \cite{nishino01, gendiar03, verstraete2004, Murg_finitePEPS_07,Cirac_iPEPS_08,Xiang_SU_08,Gu_TERG_08,Orus_CTM_09,
fu,Corboz_varopt_16, Vanderstraeten_varopt_16, Fishman_FPCTM_17, Xie_PEPScontr_17, Corboz_Eextrap_16, Corboz_FCLS_18, Rader_FCLS_18, Rams_xiD_18}. 
The infinite PEPS ansatz (iPEPS) was used to simulate unitary time evolution after a sudden Hamiltonian quench on infinite lattices \cite{CzarnikDziarmagaCorboz,HubigCirac,tJholeHubig,Abendschein08,SUlocalization,SUtimecrystal,ntu,mbl_ntu,BH2Dcorrelationspreading,ising2D_correlationsperading,schmitt2021quantum,Mazur_BH,Corboz_SF}. 
Given PEPS's non-canonical structure, it is necessary to resort to local updates in time evolution, like the neighborhood tensor update (NTU) \cite{ntu} that was used previously to simulate the many-body localization \cite{mbl_ntu}, the Kibble-Zurek ramp in the Ising and Bose-Hubbard models \cite{schmitt2021quantum,Mazur_BH,Science_Dwave}, as well as thermal states obtained by imaginary time evolution in the fermionic Hubbard model \cite{Hubbard_Sinha,Sinha_Wietek_Hubbard}. 
Here the gate sequence has the same Suzuki-Trotter structure as in the previous time evolution studies, but the gates are allowed arbitrary rotation angles instead of just small time steps. 
Expectation values in PEPS need to be evaluated in controlled-approximation schemes, like the zipper PEPS-boundary method \cite{METTS_ising_Hubbard} that we use on a finite lattice. 

The paper is organized as follows. In Sec. \ref{sec:methods} we introduce the gate sequence to be optimized. In Sec. \ref{sec:inf_1d_ising} we perform the optimization in the infinite 1D Ising quantum Ising chain. The optimal sequence is then used as a starting point for a finite chain in Sec. \ref{sec:fin_1d_ising}. The 2D model is considered in Sec. \ref{sec:2D}. We conclude in Sec. \ref{sec:conclusion}. 

\section{Gate sequence}
\label{sec:methods}

We consider the transverse-field quantum Ising model 
\be 
H = g H_1 + H_2,
\label{eq:H}
\ee 
where
\bea
H_1 &=& -\sum_i X_i, \\
H_2 &=& -\sum_{\langle i,j \rangle} Z_i Z_j,
\eea
and $Z_i=\sigma^z_i,X_i=\sigma^x_i$ are the Pauli matrices. The lattice is either a 1D chain or a 2D square lattice. 
Our goal is to obtain the ground state of the target Hamiltonian for a given value of $g$, starting from an easy-to-prepare initial state with all spins pointing along $+x$, by applying $N$ two-site $ZZ$ gates and $N$ local $X$ gates. We consider an infinite lattice first, where all rotation angles are independent of a lattice site:
\bea
U_{\textrm{BB}}\left(\beta_1,\dots,\alpha_N\right) &=& 
   e^{-i\frac12 \alpha_N g H_1} \nonumber \\
&& e^{-i\beta_N H_2}            \nonumber \\
&& e^{-i\alpha_{N-1} g H_1}     \nonumber \\
&& \dots                        \nonumber \\
&& e^{-i\alpha_1 g H_1}         \nonumber \\
&& e^{-i\beta_1  H_2}.          
\label{eq:U_BB}
\eea
Here the $2N$ angles, $\alpha_j$ and $\beta_j$, are variational parameters optimized to minimize the energy of the target Hamiltonian \eqref{eq:H}. Then the infinite-lattice optimal angles are used as a starting point for a finite lattice.

In addition to BB, we also consider adiabatic preparation (AP) as a benchmark. AP starts from the ground state of the initial Hamiltonian $g H_1$, and evolves the system to $H = g H_1+H_2$ by smoothly increasing the ferromagnetic coupling $J$ from $0$ to $1$ in finite time $\tau$: 
\bea
H &=& g H_1 + J(t) H_2, \\
J(t) &=& \frac12 + \frac12 \sin \left[ \pi \left(\frac t\tau  - \frac12\right) \right].
\eea
For the finite number of gates in \eqref{eq:U_BB}, the evolution operator can be approximated by the second-order Suzuki-Trotter decomposition, where the rotation angles are
\bea
\beta_j^{\textrm{(AP)}}    &=& \Delta t \cdot J\left[(2j-1)/(2N)\right], \nonumber \\
\alpha_{j}^{\textrm{(AP)}} &=& \Delta t.
\label{eq:alphabetaAP}
\eea 
Here the time step $\Delta t$ is the only variational parameter. To make the ramp more adiabatic, one can increase the total ramp time by increasing the time step, but a longer time step increases Suzuki-Trotter errors that excite the system above the target ground state. The optimal $\Delta t$ is a tradeoff between these two tendencies. 

\begin{figure}[t!]
\includegraphics[width=0.99\columnwidth]{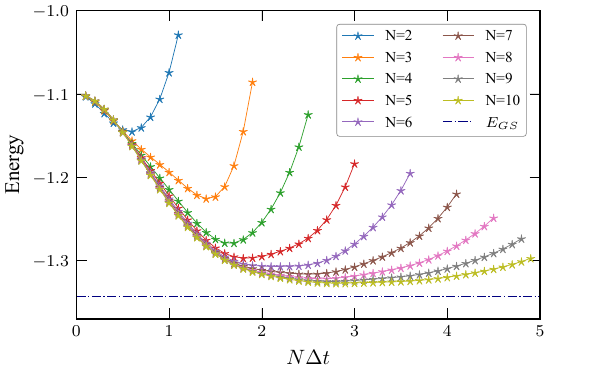}
\caption{{\bf 1D adiabatic preparation (AP).}
The final energy at the end of the AP gate sequence in function of the time step $\Delta{t}$ for different numbers of steps $N$. Here $g=1.1$ and the exact value of the ground state energy is $E_{GS}=-1.342864$. For a given $N$, increasing $\Delta t$ at first decreases the energy, by making the AP ramp slower, until the heating by Suzuki-Trotter errors prevails and the energy begins to increase.   
}
\label{fig:AP_protocol}
\end{figure}

\begin{figure}[t!]
\includegraphics[width=0.99\columnwidth]{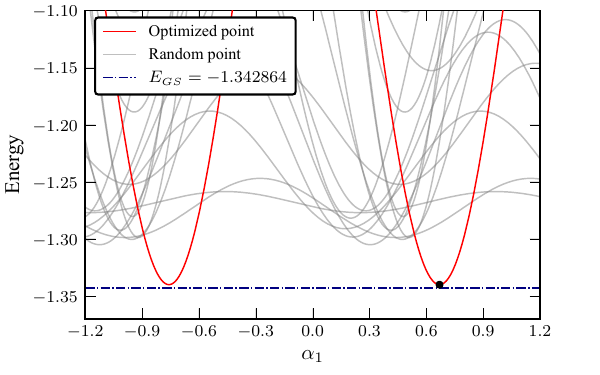}
\caption{{\bf Final energy in the BB method.}
For a randomly selected point in the $2N$-dimensional space of rotation angles, we choose one direction, say $\alpha_j$, and plot the function along this direction, keeping the other $2N-1$ parameters fixed. This cross-section is smooth with an easy-to-identify global minimum.
The final energy \eqref{eq:E_1d_inf} is periodic: if $\{ \alpha_i, \beta_j \}$ is a minimum, then $\{ \alpha_i + T_\alpha, \beta_j + T_\beta\}$ is also a minimum with the same energy. Additionally, the energy is symmetric under simultaneous sign inversion of all rotation angles: $E(\alpha_i,\beta_j)=E(-\alpha_i,-\beta_j)$. Moreover, numerical tests involving local optimization from a sample of randomly initialized points converge to the same minimum, modulo the symmetries, suggesting the existence of a unique minimum.
}
\label{fig:BB_coordinate-wise}
\end{figure}

\begin{figure}[t!]
\includegraphics[width=0.99\columnwidth]{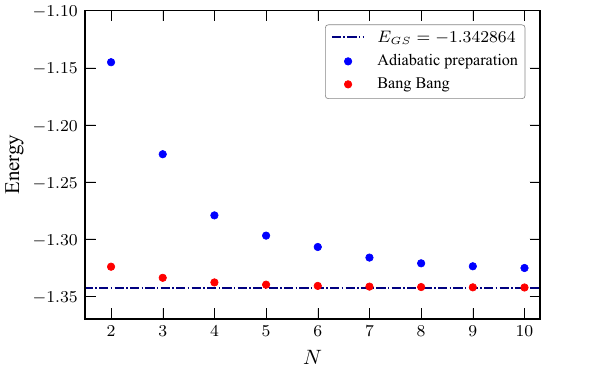}
\caption{{\bf AP vs BB in 1D.}
The optimal final energy after the $N$-gate sequence \eqref{eq:U_BB} for the AP and BB procedures. 
}
\label{fig:AP_vs_BB_energy}
\end{figure}

\begin{figure}[t!]
\includegraphics[width=0.99\columnwidth]{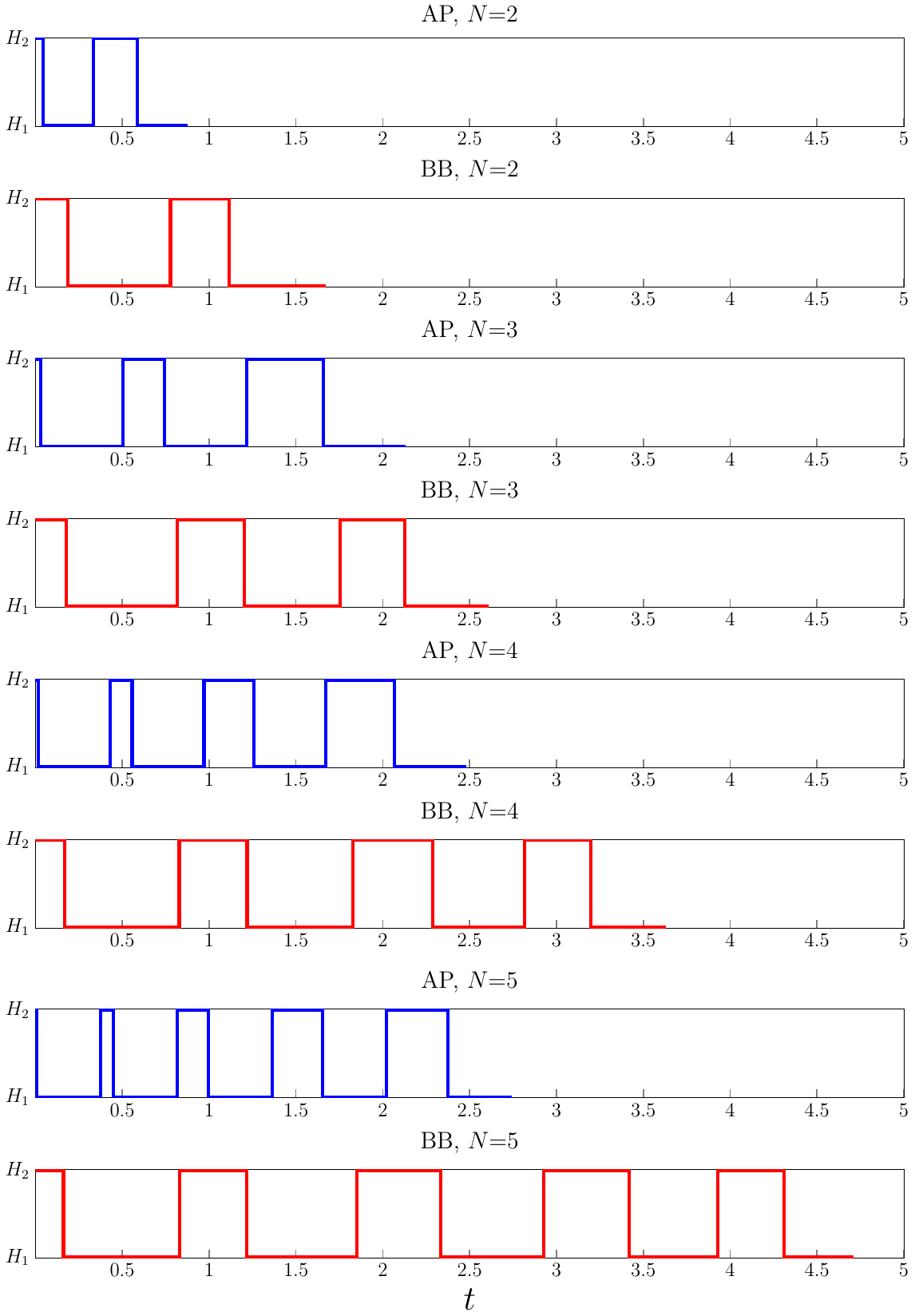}
\caption{{\bf AP vs BB sequences.}
The optimal rotation angle sequences in the AP and BB protocols for different $N$. 
The BB sequence differs non-perturbatively from the AP sequence with the same $N$.
}
\label{fig:AP_vs_BB_sequence}
\end{figure}

\begin{table}[t!]
\centering
\begin{tabularx}{0.45\textwidth} { 
  | >{\centering\arraybackslash}X| 
  | >{\centering\arraybackslash}X| 
  | >{\centering\arraybackslash}X|}
 \hline
 N & $E_\textrm{AP}$ & $E_\textrm{BB}$ \\
 \hline
 \hline
 2  & $-1.145280$ &  $-1.324301$ \\
 \hline
 3  & $-1.225828$ &  $-1.334014$\\
 \hline
 4  & $-1.279355$ &  $-1.338055$\\
 \hline
 5  & $-1.297083$ &  $-1.340026$ \\
 \hline
 6  & $-1.307025$ &  $-1.341092$ \\
 \hline
 7  & $-1.316310$ &  $-1.341711$\\
 \hline
 8  & $-1.321311$ &  $-1.342090$\\
 \hline
 9  & $-1.323919$ &  $-1.342332$\\
 \hline
 10 & $-1.325426$ & $-1.342491$ \\
 \hline
 \hline
 $E_{GS}$ exact & $-1.342864$ & $-1.342864$  \\
\hline
\end{tabularx}
\caption{
{\bf AP vs BB in 1D.}
The optimal energy after the $N$-gate sequence \eqref{eq:U_BB} for the AP and BB procedures. 
}
\label{tab:AP_vs_BB_inf1d}
\end{table}

\section{Infinite 1D model}
\label{sec:inf_1d_ising}

In this section, our aim is to obtain the best approximation of the ground state for the infinite 1D Ising model at $g=1.1$ that is close to the quantum critical point in 1D at $g=1$. At $g=1.1$ the ground state's correlation length, $\xi=10.4$, is much longer than the lattice spacing, but, can be also made much shorter than a chain of, say, $L=100$ sites. Classical simulations of the gate sequence \eqref{eq:U_BB} are performed using the iTEBD two-site canonical MPS method \cite{Orus2008} with a bond dimension $D=40$ that is sufficient to make numerical errors negligible. 

In the AP method, we optimize the time step $\Delta t$ to minimize the final energy that is plotted in Fig. \ref{fig:AP_protocol} as a function of the total evolution time $N \Delta t$. The minimization is straightforward as the energy has a single global minimum. For a given $N$, increasing $\Delta t$ at first decreases the energy by making the AP ramp slower until the Suzuki-Trotter errors prevail and the energy begins to increase. 

In the BB method, the optimization of the $2N$ rotation angles is performed using the python's \texttt{basinhopping} function from \texttt{scipy.optimize} module. It is a hybrid method that combines local optimization and random jumps, allowing us to find a global minimum.
In general, visualization of a multivariable function is a challenging task, but in the present case the function exhibits a simple form, as shown in Fig. \ref{fig:BB_coordinate-wise}. The function is smooth and its global minimum is easy to identify because its dependence on rotation angles is a multilinear polynomial, in which each variable appears with degree at most one in every term:
\bea
\mathcal{P} \left[ 
\sin(4g\alpha_i), \cos(4g\alpha_i); 
\sin(4  \beta_j), \cos(4  \beta_j)
\right].
\label{eq:E_1d_inf}
\eea
It has a period $T_\alpha=\pi/(2g)$ in the direction of $\alpha_i$, except for $\pi/g$ for $\alpha_N$, and $T_\beta=\pi/2$ for $\beta_j$.

The results are summarized in Table \ref{tab:AP_vs_BB_inf1d} and in Figs. \ref{fig:AP_vs_BB_energy} and \ref{fig:AP_vs_BB_sequence}. We can see a clear advantage of the BB algorithm over the AP one. In the next Section, we use the optimal infinite-chain BB gate sequence as a starting point for a finite chain.

\begin{figure}[h!]
\includegraphics[width=0.89\columnwidth]{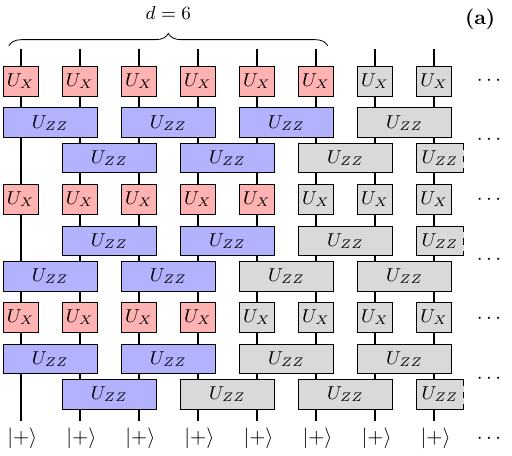}
\includegraphics[width=0.89\columnwidth]{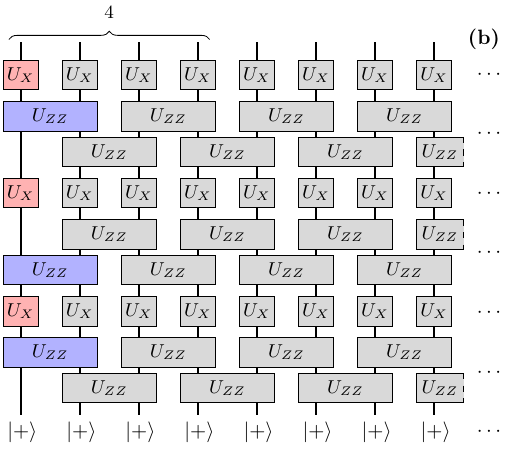}
\caption{{\bf Causal cone.}
In (a,b) the left end of a finite chain for $N=3$ bangs. 
In (a) to improve the state up to the site $d=6$, the violet and red gates should be optimized, while the gray gates can remain unchanged. This violet/red gates define a causal cone for the $d=6$ sites.
In (b) when only the end gates (violet/red) are optimized, the state is affected up to $4$ sites that are within the causal cone of the end gates. 
}\label{fig:light_cone}
\end{figure}

\begin{figure}[t!]
\includegraphics[width=0.9\columnwidth]{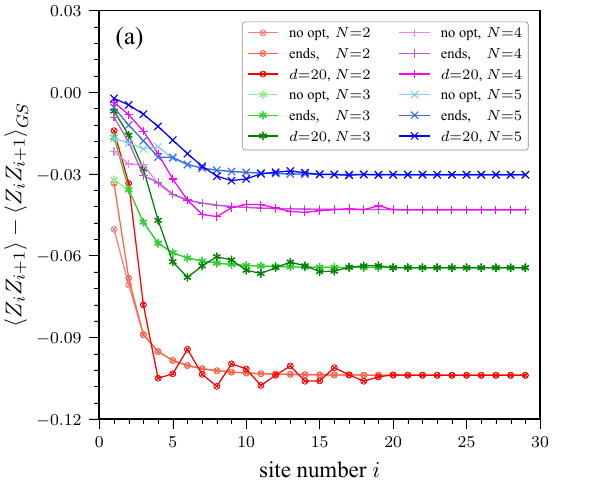}
\includegraphics[width=0.9\columnwidth]{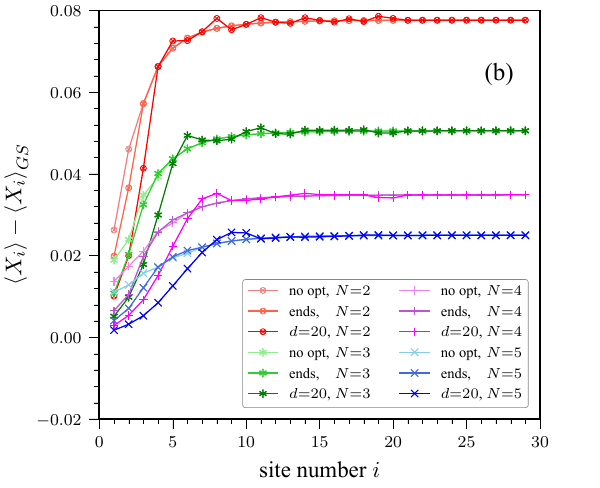}
\caption{{\bf BB for finite chain.}
BB evolution in a finite 1D chain with $L=100$ spins for $N=2,3,4,5$ bangs. The $no\; opt$ case uses rotation angles from the infinite BB for the finite-chain gate sequence. In $ends$, only the local and NN gates at the chain ends are optimized as in Fig. \ref{fig:light_cone} (b). The case $d=20$ involves full optimization of gates within the causal cone for the $d=20$ sites, which is approximately twice the correlation length $\xi=10.4$.
}
\label{fig:1D_patterns}
\end{figure}

\begin{table*}[t!]
\centering
\begin{tabularx}{1.9999\columnwidth} { 
  | >{\centering\arraybackslash}X| 
  | >{\centering\arraybackslash}X|
  | >{\centering\arraybackslash}X|
  | >{\centering\arraybackslash}X|
  | >{\centering\arraybackslash}X|}
 \hline
 N & $no\;opt$ & $ends$ & $d=10$ & $d=20$ \\
 \hline
 \hline
 2  & $-1.322036$ & $-1.322069$ & $-1.322174$ & $-1.322208$ \\
 \hline
 3  & $-1.331513$ & $-1.331543$ & $-1.331614$ & $-1.331633$ \\
 \hline
 4  & $-1.335427$ & $-1.335448$ & $-1.335494$ & $-1.335512$ \\
 \hline
 5  & $-1.337322$ & $-1.337338$ & $-1.337366$ & $-1.337385$ \\
 \hline
 \hline
 $E_{GS}$ exact & $-1.339989$ & $-1.339989$ & $-1.339989$ & $-1.339989$ \\
\hline
\end{tabularx}
\caption{
{\bf BB for finite chain.}
The total energy divided by $L=100$ sites. 
The $no\; opt$ case uses the optimal rotation angles from the infinite BB for the finite-chain gate sequence. 
In $ends$, only the local and NN gates at the chain's ends are optimized. 
The $d=10,20$ case involves full optimization of gates within the causal cone for the $d=10,20$ sites near the chain's end.
Here the correlation length $\xi=10.4$.
} 
\label{tab:1D_energies}
\end{table*}

\section{Finite 1D model}
\label{sec:fin_1d_ising}

A finite chain can be divided into its bulk and two boundaries near its ends. The optimal infinite-lattice rotation angles remain optimal in the bulk, but they are not optimal at the boundaries that extend up to the correlation/healing length distance from the ends. Therefore, optimizing the bulk gates can only yield minor gains, but optimization of the gates near the ends can improve the final result at the boundaries. In order to improve the final state to the depth $d$ from the chain's end, it is enough to optimize gates in the causal cone of this depth, see Fig. \ref{fig:light_cone}.

Fig. \ref{fig:1D_patterns} shows patterns of transverse field magnetization $\langle X_i \rangle$ and NN ferromagnetic coupling $\langle Z_iZ_{i+1}\rangle$ in function of their distance $i$ from the left end of the chain. The exact values in the ground state were subtracted from the expectation values in order to focus on the error of the BB preparation. 
The $no\; opt$ case is when the rotation angles that were optimal for the infinite chain are used everywhere on the finite one. In the bulk, i.e. more than $\xi=10.4$ sites from the end, the observables' errors are the same as in the infinite chain, as expected. Interestingly, near the end of the chain the errors are smaller, as if the BB sequence that was optimal in the bulk proved more adiabatic near the end. 

The minimal improvement is to modify only the left-most gates near the chain's end. They affect only the sites within their finite causal cone, e.g. $4$ sites for $N=3$ bangs, see Fig. \ref{fig:1D_patterns}. The same Figure shows the best pattern after optimization of all gates within the $d=20$ causal cone that is reaching to approximately twice the correlation length from the chain's end. The improvements with respect to the $no\; opt$ case do not reach far beyond the healing length $\xi=10.4$. The optimized energies are listed in Table \ref{tab:1D_energies}.

It is striking that near the end already the sequence with just $N=2$ layers gives small errors when compared to the bulk. This may be not quite surprising given that the transverse field $g=1.1$ is close to the critical $g_c=1$ in the bulk, but within the healing length from the end the system is not critical and its shorter correlations can be accurately prepared by even a shallow circuit.

\begin{figure}[t!]
\includegraphics[width=0.9\columnwidth]{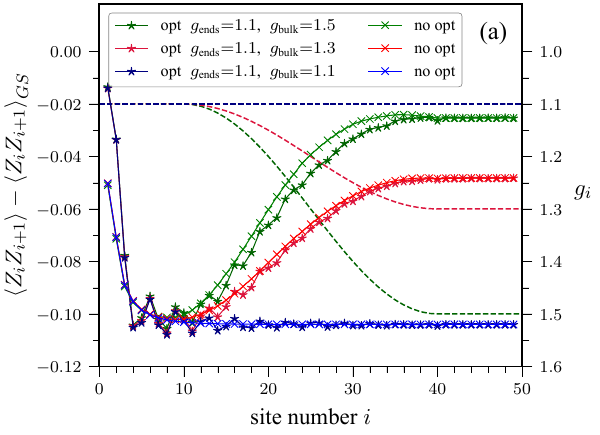}
\includegraphics[width=0.9\columnwidth]{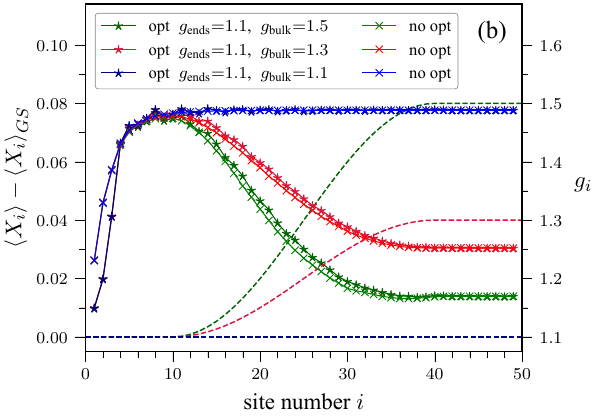}
\caption{
{\bf BB for an inhomogeneous finite chain.}
BB evolution in a finite 1D chain with $L=100$ spins for $N=2$ bangs. Transverse field $g$ (dashed lines) is set to $g_\mathrm{end}$ at the end sites $1...10$, between sites $10$ and $40$ it increases sinusoidally to the bulk value $g_\mathrm{bulk}$, and remains at $g_\mathrm{bulk}$ in the bulk region. The BB evolution is performed using position-dependent rotation angles, which are obtained from optimizing infinite chain for a local value of $g$. Further optimization does not make much difference.
}
\label{fig:1D_critical_ends}
\end{figure}

Turning the argument around, in Fig. \ref{fig:1D_critical_ends} we consider a finite chain with a site-dependent transverse magnetic field $g_i$. It is near-critical close to the end of the chain, $g_\mathrm{end}=1.1$, and then smoothly ramps over $30$ sites up to a higher bulk value $g_\mathrm{bulk}=1.1,1.3,1.5$. As the ramp is wider than the maximal local healing length, equal to $\xi\approx10$ at $g_\mathrm{end}$, it is justified to make an approximation where local rotation angles at site $i$ are set equal to the rotation angles optimized on an infinite chain with a uniform transverse field equal to $g_i$. 

\begin{figure}[t!]
\includegraphics[width=5.5cm]{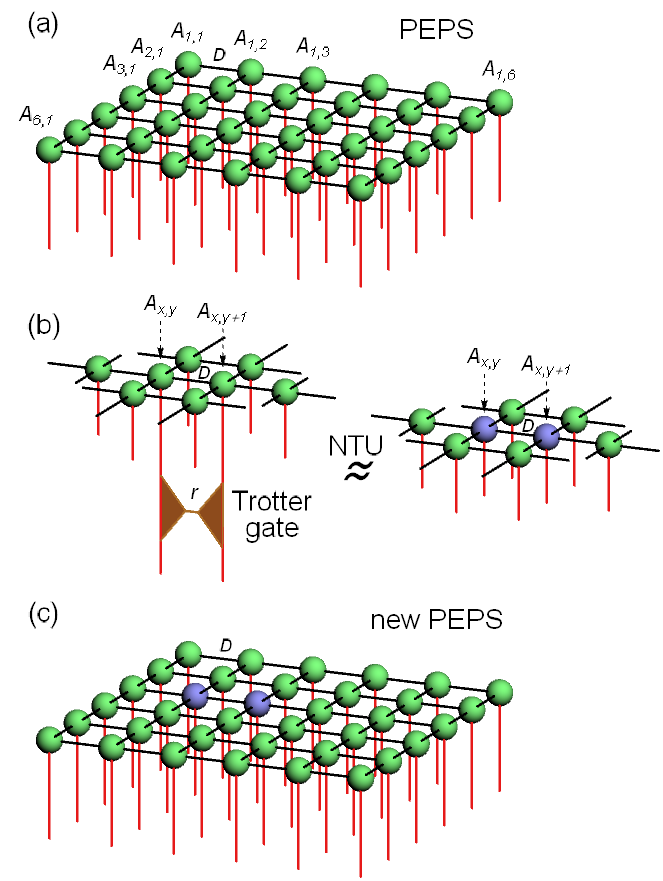} 
\figcaption{
{\bf NTU. }
In (a) infinite PEPS with tensors $A$ (purple) and $B$ (pink) on the two sub-lattices of an infinite checkerboard lattice. The red lines are physical spin indices and the black lines are bond indices, with bond dimension D, contracting NN site tensors. 
In one of Suzuki-Trotter steps a Trotter gate is applied to every NN pair of $A$-$B$ tensors along every horizontal row (but not to horizontal $B$-$A$ pairs). The gate can be represented by a contraction of two tensors, $G_A$ and $G_B$, by an index with dimension $r$. When the two tensors are absorbed into tensors $A$ and $B$ the bond dimension between them increases from $D$ to $rD$.
In (b) the $A$-$B$ pair -- with a Trotter gate applied to it -- is approximated by a pair of new tensors, $A'$ (deep purple) and $B'$ (darker blue), connected by an index with the original dimension $D$. The new tensors are optimized to minimize the difference between the two networks in (b).
After $A'$ and $B'$ are converged, they replace all tensors $A$ and $B$ in a new iPEPS shown in (c). 
Now the next Trotter gate can be applied.
}
\label{fig:NTU}
\end{figure}

\section{2D model}
\label{sec:2D}

The infinite 2D model was considered in the previous work \cite{BB_iPEPS} at a transverse field $g=3.1$ that is close to the quantum critical point at $g_c=3.04438(2)$~\cite{Deng_QIshc_02}. With the help of the variational algorithm for the 2D iPEPS tensor network \cite{corboz16b} the correlation length in the ground state was estimated as $\xi=2.38$. 

In Ref. \cite{BB_iPEPS} the target energy after the 2D BB sequence \eqref{eq:U_BB} was obtained by evolving iPEPS with the neighborhood tensor update (NTU) algorithm \cite{ntu}, see Fig. \ref{fig:NTU}. The algorithm involves an approximation: the bond dimension increased by application of a NN Trotter gate has to be truncated back to its original value, see Fig. \ref{fig:NTU} (b). 
For each NN gate, the Frobenius norm of the difference between the left ($L$) and right ($R$) hand sides of Fig. \ref{fig:NTU} (b) is minimized. The NTU error $\delta_{i}$ of the $i^{\textrm{th}}$ gate is defined as the minimal norm $||L-R||$. $\delta_i$ is a rough estimate for an error inflicted on local observables by the bond dimension truncation. Accumulating truncation errors can eventually derail the time evolution. In the worst-case scenario, the errors are additive. This motivates a total NTU error \cite{Hubbard_Sinha}: 
\be 
\epsilon_{\textrm{NTU}} = \sum_i \delta_{i},
\label{eq:NTUerr}
\ee 
where the sum is over all performed NN Trotter gates. In Ref. \cite{BB_iPEPS} the total error was kept at the level of $10^{-8}$ to warrant that the gate sequence simulated by the NTU evolution gives the same result as its implementation with quantum gates.

In Ref. \cite{BB_iPEPS} BB protocols with $N=2..6$ were compared with a Suzuki-Trotter decomposition of an adiabatic quantum state preparation (AP). The BB protocol was able to obtain much lower energy in a smaller number of steps. 
The optimal BB angles could have ended up being close to their AP values, with small perturbations meant to approximate the counter-adiabatic term \cite{del_Campo_2019} by the Suzuki-Trotter errors \cite{Wurtz2022counterdiabaticity}, or to follow a qualitatively different path along a non-perturbative short-cut to the desired ground state. The latter turned out to be the case.

\begin{figure}[t!]
\centering
\includegraphics[width=0.99\columnwidth]{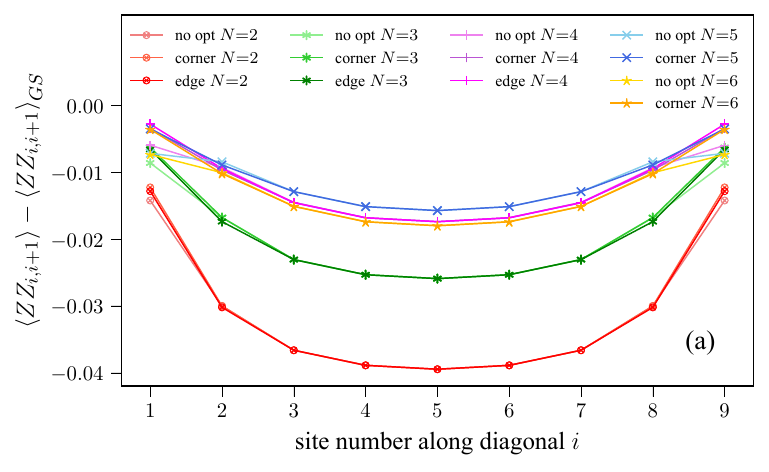}
\includegraphics[width=0.99\columnwidth]{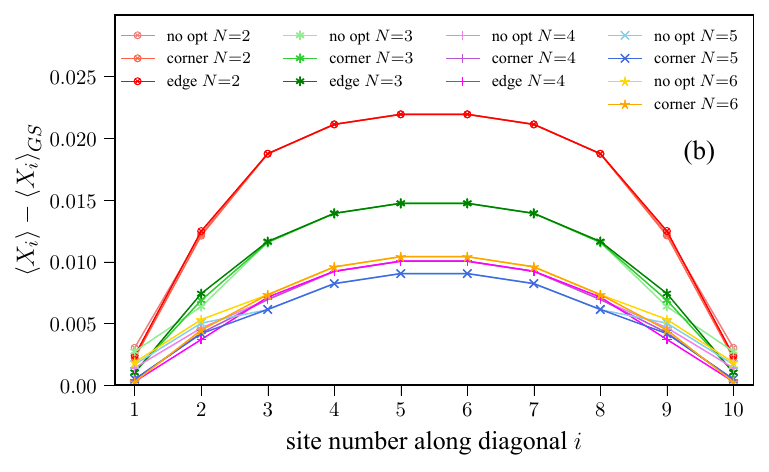}
\caption{{\bf BB for a finite 2D lattice.} 
BB evolution in a $10 \times 10$ 2D Ising model for $N = 2, 3, 4, 5, 6$ bangs at $h = 3.1$. The calculations are performed with bond dimension $D = 6$ and $\chi = 40$. 
The $no\; opt$ case uses rotation angles from an infinite BB evolution applied to a finite system. 
In the $corner$ case, local and nearest-neighbor gates at the 4 corners (4 local gates and 8 NN gates) are optimized. Due to symmetry between the 4 corners, only one $\alpha$ and one $\beta$ parameter are optimized for each bang.
In the $edge$ case, local and NN gates at the 4 edges are optimized.
}
\label{fig:2D_patterns}
\end{figure}

\begin{table*}[t!]
\centering
\begin{tabularx}{1.9999\columnwidth} { 
  | >{\centering\arraybackslash}X| 
  | >{\centering\arraybackslash}X|
  | >{\centering\arraybackslash}X|
  | >{\centering\arraybackslash}X|
  | >{\centering\arraybackslash}X|}
 \hline
 N & Trunc. err. & $no\;opt$ & $corners$ & $edges$ \\
 \hline
 \hline
 2 & $8.32\times10^{-16}$ & $-3.255562$ & $-3.255607$ & $-3.255676$ \\
 \hline
 3 & $7.29\times10^{-5}$ & $-3.259049$ & $-3.259083$ & $-3.259124$ \\
 \hline
 4 & $4.75\times10^{-4}$ & $-3.260810$ & $-3.260838$ & $-3.260876$ \\
 \hline
 5 & $8.78\times10^{-4}$ & $-3.260890$ & $-3.260918$ & $-$ \\
 \hline
 6 & $4.65\times10^{-4}$ & $-3.260742$ & $-3.260772$ & $-$ \\
 \hline
 \hline
 $E_{GS}$ ITE & $-$ & $-3.262603$ & $-3.262603$ & $-3.262603$ \\
\hline
\end{tabularx}
\caption{
{\bf BB for a finite 2D lattice.}
The total energy divided by $L^2$ sites. 
The $no\; opt$ case uses the optimal rotation angles from the infinite BB for the finite system gate sequence. 
In $corners$, only the local and NN gates at the corners are optimized. 
The $edges$ case involves full optimization of gates on the edge of the grid.
Here the correlation length $\xi=2.38$ and the bond dimensions $D=6$ and $\chi = 40$.
} 
\label{tab:2D_energies}
\end{table*}

We use the infinite sequence obtained in Ref. \cite{BB_iPEPS} for $g=3.1$ as a starting point for a $L\times L$ open square lattice with $L=10$ and the same $g$. The reference ground state on the finite lattice is obtained by imaginary time evolution with the NTU algorithm and expectation values are evaluated by the zipper method \cite{METTS_ising_Hubbard} with a bond dimension $\chi$ of the PEPS boundary. All numerical procedures available in the YASTN package \cite{YASTN1,YASTN2}. 

Figure \ref{fig:2D_patterns} shows errors of the transverse magnetization and the NN ferromagnetic correlators after embedding the infinite BB sequence in the finite lattice. 
They are compared with the data when only corner gates or the gates along the edges of the lattice are further optimized. 
Table \ref{tab:2D_energies} collects the energies of the states before and after the optimization. 
As in 1D, the optimization brings some improvement to the errors that are already small at the boundaries. It affects the state within the causal cone of the optimized gates.  

\section{Conclusion}
\label{sec:conclusion}

The earlier paper \cite{BB_iPEPS} demonstrated that the infinite PEPS can be used to optimize shallow bang-bang quantum gate sequences preparing the ground state of a 2D quantum Hamiltonian in an infinite system. In this work we use the infinite-lattice BB sequence as a starting point to optimize BB sequences on a finite lattice. As it is already optimal in the bulk, the optimization needs to be performed only in the causal cone of the lattice boundary defined as sites within the healing length from the edges.
With the bulk close to quantum criticality, even without any optimization, the errors at the boundary are lower than in the bulk. The boundary is less critical and the ground state there can be better approximated with a small number of bangs. 
Furthermore, for a Hamiltonian varying in space slowly compared to the local correlation length, an accurate BB sequence near a given site can be obtained by optimizing the sequence on an infinite lattice with the local Hamiltonian near the site.

The figure data can be downloaded from \url{https://uj.rodbuk.pl/dataset.xhtml?persistentId=doi:10.57903/UJ/30GXBQ} 

\acknowledgments
%
We are indebted to Piotr Czarnik for a discussion.
This research was supported in part by the National Science Centre (NCN), Poland under project 2019/35/B/ST3/01028 (I.S.) and
project 2021/03/Y/ST2/00184 within the QuantERA II Programme that has received funding from the European Union Horizon 2020 research and innovation programme under Grant Agreement No 101017733 (J.D.).
The research was also supported by a grant from the Priority Research Area DigiWorld under the Strategic Programme Excellence Initiative at Jagiellonian University.

\bibliography{KZref.bib} 

\end{document}